# Inverse Faraday effect and Stokes drift in plasma

A. Y. Bekshaev

Physics Research Institute, I.I. Mechnikov National University, Odessa, Ukraine;
*bekshaev@onu.edu.ua

**Abstract**

Recent theory of the light-induced medium magnetization (inverse Faraday effect, IFE) performed by a transversely-limited circularly-polarized light beam [Phys. Rev. B **91**, 020411 (2015)] predicts the existence of a "demagnetization current" (DC) at the beam periphery which, apparently, acts oppositely to the light-induced rotational motion of the charge carriers inside the beam and thus reduces the IFE by the factor of 2. In this note, taking the longitudinal component of the beam into account, we show that the peripheral DC is two times higher than was calculated before. Nevertheless, this circumstance does not cancel the IFE because the DC, as a sort of Stokes-drift current in plasma [Phys. Rev. E **105**, 065208 (2022)], is accompanied by the additional "magnetization current" of the opposite direction.

When a circularly polarized (CP) light beam propagates through a medium with free charges (electrons), these are set in a rotational motion which induces the medium magnetization; this is the essence of "inverse Faraday effect" (IFE) [1–5]. Authors of the paper [3] have shown that in the usual case of spatially limited illuminating beam, the IFE is accompanied by the macroscopic drift of electrons circulating around the beam center, which produces a circular "demagnetization current" essentially decreasing the IFE-generated magnetic momentum induced in the medium. Now we readdress this problem based on the recently developed general approach to electromagnetic momentum and wave-induced drift phenomena in plasma [6,7].

Following to [3], we consider a monochromatic CP beam with frequency $\omega$ and wavenumber $k = \omega/c$ ($c$ is the light velocity) propagating in a homogeneous medium along axis $z$; the medium is conductive and contains free electrons of the charge $e < 0$ and mass $m$ with the volume density $n_e$ (in the equilibrium conditions, the charge of electrons is balanced by the "background" positive charge of unmovable ions). Let the electric vector transverse components of the beam field be

$$E_x = E_\perp, \ E_y = i\sigma E_\perp \qquad (1)$$

where $\sigma = \pm 1$ is the CP handedness, and $E_\perp \equiv E_\perp(x, y)$ is the Gaussian function of transverse coordinates,

$$E_\perp(x, y) = E_{\perp 0} \exp\left(-\frac{r^2}{w^2}\right) \qquad (2)$$

where $r = \sqrt{x^2 + y^2}$ is the polar radius within the beam cross section. As usual, we employ the complex representation of harmonic functions in which their real instantaneous values, e.g., $E_x(t)$, are determined as $E_x(t) = \text{Re}\left[E_x \exp(-i\omega t)\right]$, etc. In this representation, the electron velocities are related with the local values of the electric-field components according to equations [6]

$$\mathbf{v} = \frac{ie}{m\omega}\mathbf{E}; \qquad (3)$$

in particular,

$$v_x = \frac{ie}{m\omega}E_x = \frac{ie}{m\omega}E_\perp, \quad v_y = \frac{ie}{m\omega}E_y = -\sigma\frac{e}{m\omega}E_\perp, \tag{4}$$

and the electrons perform circular motion with the orbit radius $a = \frac{e}{m\omega^2}E_\perp$. In [3], the consideration is restricted to the transverse components but, in fact, any transversely limited beam contains also the longitudinal electric-field component [8–12]:

$$\mathbf{E} = \mathbf{e}_x E_x + \mathbf{e}_y E_y + \mathbf{e}_z E_z \tag{5}$$

where $\mathbf{e}_x$, $\mathbf{e}_y$, $\mathbf{e}_z$ are the unit vectors of the coordinate axes, and

$$E_z = \frac{i}{k}\left(\frac{\partial E_x}{\partial x} + \frac{\partial E_y}{\partial y}\right) \sim \gamma E_\perp, \tag{6}$$

$\gamma = (kw)^{-1}$ is the small parameter of the paraxial approximation [8–10]. Note that in paraxial fields all the field components show an oscillatory behavior along the longitudinal direction which is expressed by the factor $\exp(ikz)$ implicitly entering each term of equations (2) – (6).

For determinacy, our further analysis is localized to the initial cross section $z = 0$ (and, maybe, its nearest vicinity, see below), which is supposed to coincide with beam waist. The rotational motion of electrons, described by Eqs. (4), is the source of the material electromagnetic spin $\mathbf{s}^M$ and magnetization $\mathbf{M}$ of the medium whose densities are determined by Eqs. (43) and (44) of Ref. [6]:

$$\mathbf{s}^M = \frac{mn_e}{2\omega}\mathrm{Im}(\mathbf{v}^* \times \mathbf{v}), \quad \mathbf{M} = \frac{e}{2mc}\mathbf{s}^M. \tag{7}$$

With account for Eqs. (2), (3) and (4), this results in

$$\mathbf{M} = \frac{e}{16\pi mc\omega}\frac{\omega_p^2}{\omega^2}\mathrm{Im}(\mathbf{E}^* \times \mathbf{E}), \quad \omega_p^2 = \frac{4\pi e^2 n_e}{m} \tag{8}$$

($\omega_p$ is the plasma frequency), which exactly coincides with the magnetization calculated in Eq. (11) of [3] – the only precaution is to convert the expression from Gaussian to SI units, which can be made via replacements [13]

$$\mathbf{M} \to \mathbf{M}\sqrt{\frac{\mu_0}{4\pi}}, \quad e \to \frac{e}{\sqrt{4\pi\varepsilon_0}}, \quad \mathbf{E} \to \mathbf{E}\sqrt{4\pi\varepsilon_0}, \quad c \to \frac{1}{\sqrt{\varepsilon_0\mu_0}}.$$

Taking only the transverse electric field components $E_x$, $E_y$, expression (8) reduces to the form

$$\mathbf{M} = \mathbf{e}_z \frac{\sigma e^3 n_e}{2m^2 c\omega^3} E_{\perp 0}^2 \exp\left(-\frac{2r^2}{w^2}\right). \tag{9}$$

which describes the magnetization density in the plasma illuminated by the Gaussian CP beam (1), (2). The allowance for longitudinal velocity components following from (3) and (5) give small corrections of (9) which are neglected in further reasonings.

The paper [3] indicates that the microscopic circulation of electrons described by Eqs. (4) is not the only source of the medium magnetization. The results (8), (9) are obtained upon assumption that the electric field amplitude is the same on the whole electron's trajectory, which is only correct if the electron's orbit radius is infinitely small or the optical field is spatially homogeneous. In real situations, different parts of the electron's orbit "feel" different electric field strengths, according to (2), and this invokes a macroscopic regular drift of electrons. The drift velocity is calculated in [3] by expanding the instantaneous electron velocity in a perturbation series, where the high oscillatory velocity is supplemented by small "slow" drift components, and the time-averaging over the optical oscillation period is performed. Here, we consider the drift behavior of the oscillating electrons by means of the recently proposed regular way based on the Stokes drift concept [6,7]. According to this concept, in a wave field, the charge carriers (electrons) experience a regular "slow" drift whose velocity can be determined as

$$\mathbf{v}_d = \frac{1}{2\omega} \operatorname{Im}\left[\left(\mathbf{v}^* \cdot \nabla\right)\mathbf{v}\right]. \tag{10}$$

For a paraxial field, since $(\partial/\partial z)[...] \simeq ik[...]$ and, according to (6), $v_z = \frac{i}{k}\left(\frac{\partial v_x}{\partial x} + \frac{\partial v_y}{\partial y}\right)$, Eq. (10) leads to expressions

$$\mathbf{v}_d = \frac{1}{2\omega}\operatorname{Im}\left[\left(v_x^*\frac{\partial}{\partial x} + v_y^*\frac{\partial}{\partial y} + v_z^* \cdot ik\right)\left(\mathbf{e}_x v_x + \mathbf{e}_y v_y + \mathbf{e}_z v_z\right)\right] = \mathbf{v}_{d\perp} + \mathbf{v}_{d\|}; \tag{11}$$

$$\mathbf{v}_{d\perp} = \frac{1}{2\omega}\operatorname{Im}\left[\left(v_x^*\frac{\partial}{\partial x} + v_y^*\frac{\partial}{\partial y}\right)\left(\mathbf{e}_x v_x + \mathbf{e}_y v_y\right)\right], \quad \mathbf{v}_{d\|} = \frac{1}{2\omega}\operatorname{Im}\left[\left(\frac{\partial v_x^*}{\partial x} + \frac{\partial v_y^*}{\partial y}\right)\left(\mathbf{e}_x v_x + \mathbf{e}_y v_y\right)\right] \tag{12}$$

(only terms of the first order in $\gamma$ (6) are kept).

These results show that the drift velocity consists of two summands. The first summand $\mathbf{v}_{d\perp}$ follows from the transverse field components only, the second one $\mathbf{v}_{d\|}$ appears due to the longitudinal field (6). Under conditions (4), the expression (12) for $\mathbf{v}_{d\perp}$ can be represented as

$$\mathbf{v}_{d\perp} = \frac{\sigma e^2}{2m^2\omega^3} E_\perp \left(\mathbf{e}_y \frac{\partial}{\partial x} - \mathbf{e}_x \frac{\partial}{\partial y}\right)E_\perp = \mathbf{e}_\phi \frac{\sigma e^2}{2m^2\omega^3} E_\perp \frac{\partial E_\perp}{\partial r} \tag{13}$$

where $\mathbf{e}_\phi$ is the unit vector of the polar azimuth $\phi = \arctan(y/x)$. With this drift component, the electric current is associated

$$\mathbf{j}_{d\perp} = en_e \mathbf{v}_{d\perp} \tag{14}$$

which exactly coincides with the drift current calculated earlier (see Eq. (24) in [3]). Therefore, the Stokes drift in plasma explains the drift current phenomenon described in [3] and confirms its numerical value. However, our analysis also reveals that the result of [3], based on the transverse optical-field components, is incomplete; there exists an additional drift-current part emerging due to the longitudinal field component and associated with the term $\mathbf{v}_{d\|}$ of Eqs. (11), (12). Based on Eqs. (4) and (12), one can easily find that

$$\mathbf{v}_{d\|} = \mathbf{e}_\phi \frac{\sigma e^2}{2m^2\omega^3} E_\perp \frac{\partial E_\perp}{\partial r} = \mathbf{v}_{d\perp} \tag{15}$$

and, therefore, that the total "macroscopic drift current" accompanying the IFE is two times higher than the value obtained in [3]:

$$\mathbf{j}_d = \mathbf{j}_{d\perp} + \mathbf{j}_{d\|} = 2\mathbf{j}_{d\perp} = \mathbf{e}_\phi \frac{\sigma e^3 n_e}{m^2\omega^3} E_\perp \frac{\partial E_\perp}{\partial r} = -\mathbf{e}_\phi \frac{2\sigma e^3 n_e}{m^2\omega^3} E_{\perp 0}^2 \frac{r}{w^2}\exp\left(-\frac{2r^2}{w^2}\right). \tag{16}$$

[Note that the longitudinal-field contribution $\mathbf{j}_{d\|}$ can be, of course, calculated by the perturbation method accepted in [3]. To this end, Eq. (15) of [3] should be supplemented by the term accounting for the field variation along axis $z$:

$$\mathbf{E}(y,z) = \mathbf{E}_{gc} + (y - y_{gc})\frac{\partial E_\perp}{\partial y}\frac{\mathbf{E}_{gc}}{E_\perp} + (z - z_{gc})\frac{\partial E_\perp}{\partial z}\frac{\mathbf{E}_{gc}}{E_\perp}$$

where the subscript "$gc$" denotes the quantities related to the "gyration center", and

$$z - z_{gc} \simeq \frac{1}{\omega}v_z = -\frac{e}{m\omega k}\left(\frac{\partial E_\perp}{\partial x} + i\sigma\frac{\partial E_\perp}{\partial y}\right), \quad \frac{\partial E_\perp}{\partial z} \simeq ikE_\perp.$$

After these transformations, repeating the procedure of Eqs. (16) – (24) of [3] and properly time-averaging directly leads to our results (15) and (16)].

The correction (16) of the previous finding of [3] looks just quantitative but actually it imposes rather dramatic conclusions. Really, the drift current (14) was interpreted [3] as an additional source of magnetization which acts oppositely to the microscopic orbiting of individual electrons, and its contribution decreased the total magnetic moment of the medium by a factor of two (see Eqs. (13) and (28) of [3]). Consequently, the presence of yet another summand $\mathbf{j}_{d\parallel}$ of the same value would completely eliminate the IFE magnetization! This conclusion looks quite paradoxical, especially in view of the well known experimental confirmations of the IFE (see, e.g., [5]) and its phenomenological explanations [1,4].

To resolve this paradox, one may notice that the drift current, introduced in [3] and analyzed in the above paragraphs of the present note, is not the only macroscopic current induced by the electromagnetic field in plasma. Really, the material spin (7) is accompanied by the spin momentum [6,9,10,14–16] with the density $\mathbf{p}_S^M = \frac{1}{2}\nabla \times \mathbf{s}^M$, which implies the corresponding electric current

$$\mathbf{j}_S^M = \frac{e}{m}\mathbf{p}_S^M = c\nabla \times \mathbf{M}. \tag{17}$$

Remarkably, this is just the "magnetization current" associated with the magnetic moment density $\mathbf{M}$ [1,17,18] (in this context, the drift current (16) can be called "demagnetization current"). Using expression (9), one can easily find its explicit form for the field configuration analyzed in this note:

$$\mathbf{j}_S^M = \mathbf{e}_\phi \frac{2\sigma e^3 n_e}{m^2 \omega^3} E_{\perp 0}^2 \frac{r}{w^2} \exp\left(-\frac{2r^2}{w^2}\right), \tag{18}$$

which is exactly the opposite of the drift current (16). This rather interesting result is, in fact, quite expectable because it demonstrates a special case of the mutual compensation of the Stokes-drift and spin-momentum contributions substantiated in the general form by Eqs. (41), (42) and (44), (47) of Ref. [6].

To sum up, we emphasize that, contrary to the conclusion of Ref. [3], the "demagnetization current" (16) does not generate the medium magnetization opposite to the IFE magnetization (9) (and accordingly, does not diminish or nullify the IFE itself). Actually, the current (16) is compensated by the "magnetization current" (18) and is thus not able to affect the magnetization induced by the microscopic orbital motion of electrons. Moreover, the IFE with spatially limited magnetizing beams is not coupled with any "visible" macroscopic current circulating around the perimeter of the laser beam (at least, in the simple medium and beam configurations considered here and in [3]). At the same time, there are important differences in physical nature of the mutually compensating currents (16) and (17), (18). The drift current (16) is coupled with the real azimuthal flux of charges while the current (17), (18) is "bound" [17,18]: it performs no real charge transport and cannot be measured by an ammeter. It expresses the combined action of the microscopic orbital motions of all electrons and can be interpreted as a result of incomplete compensation of the linear velocity components of the adjacent microscopic electron orbits [9,15,19].

Undoubtedly, the same mechanism of the macroscopic current formation is qualitatively applicable to the IFE generated by any spatially-limited light beam. For example, consider a flat-top beam with an arbitrary asymmetric intensity profile (Fig. 1). Inside the central part of the beam cross section (enclosed by the contour A), the intensity is assumed homogeneous, so that the electrons describe absolutely identical and perfectly circular orbits thus producing a homogeneous magnetization. Simultaneously, the microscopic motions of adjacent electrons completely cancel each other, which leaves no room for macroscopic currents. However, at the cross-section periphery (between the contours A and B), the beam intensity falls down; the electric-field is distributed inhomogeneously, which is the source of the drift current $\mathbf{j}_d = en_e\mathbf{v}_d$ (10) – (12) and of the magnetization current $\mathbf{j}_S^M$ (17). However, the resulting macroscopic current is not observable because, generally, $\mathbf{j}_d = -\mathbf{j}_S^M$ [6], and, anyway, it does not affect the IFE-induced magnetization of the illuminated medium region (interior of the contour B).

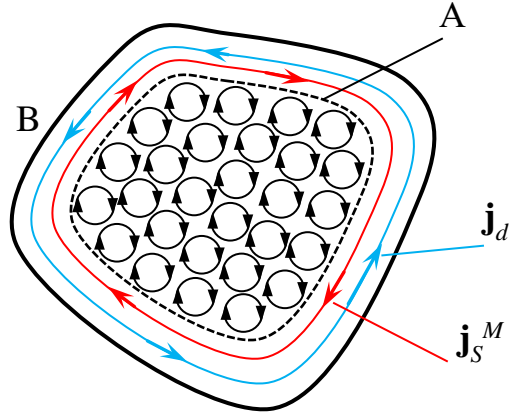

Fig. 1. Cross section of a flat-top beam. The light intensity is constant within the central area enclosed by the dashed contour A and decreases to zero in the space between the contour A and the solid contour B (the beam "boundary"). Blue and red contours with arrows schematically indicate the drift current $\mathbf{j}_d$ (16) and the magnetization current $\mathbf{j}_S^M$ (17).

In can be noted, however, that the full compensation of the drift current $\mathbf{j}_d$ and the magnetization current $\mathbf{j}_S^M$ does not mean the full compensation of other, non-electromagnetic aspects of the Stokes drift (10) and the material spin momentum $\mathbf{p}_S^M$ (see (17)). This follows from the difference in their physical nature [6,14,15]. In particular, the real transfer of mass and charge associated with the Stokes-drift momentum $\mathbf{p}_d = (m/e)\mathbf{j}_d$ can produce a specific mechanical action that presumably can be detected if the plasma medium contains suspended (embedded) particles, impurity atoms, etc. On the contrary, the spin momentum performs no mass transfer (for which reason it is called sometimes "virtual" [20]), and its mechanical manifestation, if it exists, will look quite differently. Such expectations are supported by the fact that the drift momentum $\mathbf{p}_d$ is a part of the "orbital" (canonical) momentum while the material spin momentum $\mathbf{p}_S^M$ belongs to the spin part of the spin-orbital (canonical) electromagnetic-momentum decomposition [9,10,16,20,21], and the difference in mechanical actions of the spin and orbital momenta in vacuum are well established [20,22].